\documentclass[twocolumn,12pt,cites,graphicx]{article}
\usepackage{graphicx}
\usepackage{epsfig}
\title{Curvature Distribution of Worm-like Chains in Two and Three Dimensions}
\author{Shay M. Rappaport, Shlomi Medalion and Yitzhak Rabin\thanks{Nano-materials Research Center, Institute ofNanotechnology and Advanced Materials, Bar-Ilan University, Ramat-Gan 52900,} \\Department of Physics, Bar-Ilan University, Ramat-Gan 52900, Israel}
\begin{document}

\maketitle
\renewcommand{\thefootnote}{\fnsymbol{footnote}}

\noindent Bending of worm-like polymers carries an energy penalty which results in the
appearance of a persistence length $l_p$ such that the polymer is straight
on length scales smaller than $l_p$ and bends only on length scales larger
than this length. Intuitively, this leads us to expect that the most
probable value of the local curvature of a worm-like polymer undergoing
thermal fluctuations in a solvent, is zero. We use simple geometric
arguments and Monte Carlo simulations to show that while this expectation is
indeed true for polymers on surfaces (in two dimensions), in three
dimensions the probability of observing zero curvature anywhere along the
worm-like chain, vanishes.
\section{Introduction}
\label{intro}
Semi-flexible polymers are usually modeled as space curves (either
continuous or discrete) that possess bending rigidity. Following the linear
theory of elasticity of slender rods, the bending energy of such worm-like
polymers is taken to be a quadratic functional of the local curvature \cite
{Landau-Lifshitz}. The curvature is a local geometrical property of a
continuous curve - the reciprocal of the radius of the tangential circle at
any point along the curve. In discrete models the curvature is the angle
(per unit length) between adjacent segments, in the limit when the length of
these segments goes to zero (see below). The calculation of the statistical
properties of such polymers involves generating a representative statistical
ensemble of different conformations of space curves and weighting them by
appropriate Boltzmann factors. Since these statistical weights are Gaussian
functions of the local curvature, one expects that the most probable value
of the curvature is zero, i.e., that the probability distribution of
curvature is peaked at the origin. This expectation agrees with our
intuition about semi-flexible polymers: since they deviate from straight
lines only on length scales exceeding their persistence length, the local
curvature of a typical configuration is expected to vanish almost
everywhere. As we will show in the following using both analytical arguments
and Monte Carlo simulations, the above conclusions hold only in two
dimensions; surprisingly, the most probable value of the curvature in three
dimensions is not zero!
\section{Curves, curvature and measure}
\begin{figure}[ht]
\begin{center}
\includegraphics{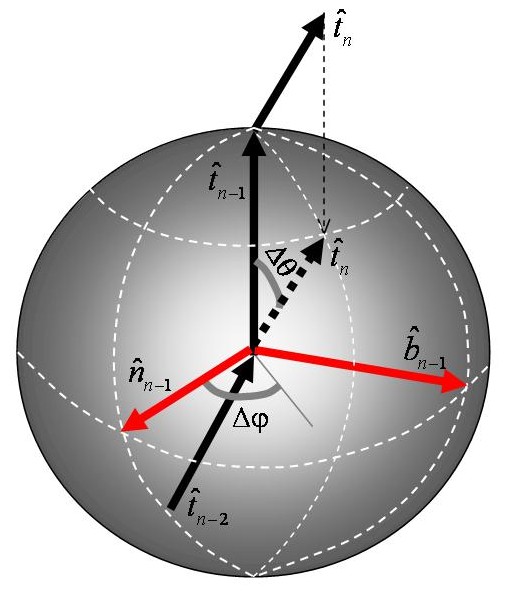}
\caption{Caption of Fig. 1.}
\label{fig:tangent}
\end{center}
\end{figure}

Consider a space curve defined by a vector $\vec{r}(s)$ (with respect to
some space-fixed coordinate system), where the parameter $s$ is the
arclength. The shape and the orientation of the curve are completely
determined (up to unifrom translation) by the direction of the tangent to
the curve at every point along its contour, $\hat{t}(s)=d\vec{r}/ds$:
\begin{equation}
\vec{r}(s)=\int_0^sds^{\prime}\hat{t}(s^{\prime})
\end{equation}
Since the tangent is a unit vector, $|\hat{t}(s)|=1$, it can be represented
as a point on a sphere of unit radius; the coordinates of this point are
given by the two spherical angles {$\theta$ and $\phi$}. Finally, one can
introduce a local property of the curve, the curvature, defined as :
\begin{equation}
\kappa(s)=\left|\frac{d^2\vec{r}}{ds^2}\right|=\left|\frac{d\hat{t}}{ds}
\right|  \label{eq:kappa}
\end{equation}
Let us introduce a discrete version of the continuous curve of length $L$,
by dividing it into $N$ segments of length $\Delta s$ ($N\Delta s=L$), such
that the orientation of the n'th segment connecting points $s$ and $s+\Delta
s$ is defined by the direction of the tangent at point $s$. One can define
the orientation of the $n$'th tangent vector (i.e., the $n$'th segment) in
terms of the ($n-1$)'th one
\begin{equation}
\begin{array}{c}
\hat{t}_n=\cos(\Delta\theta)\hat{t}_{n-1}+\cos(\Delta\varphi)\sin(\Delta
\theta)\hat{n}_{n-1} \\
+\sin(\Delta\varphi)\sin(\Delta\theta)\hat{b}_{n-1}
\end{array}
\label{eq:tangent}
\end{equation}
As shown in Fig. \ref{fig:tangent}, $\Delta\theta\in[0..\pi]$ is the angle
between the $n$'th and the ($n-1$)'th segments and $\Delta\varphi\in[
-\pi..\pi]$ is the polar angle between the two planes defined by $\hat{t}
_{n-2}\times\hat{t}_{n-1}$ and $\hat{t}_{n-1}\times\hat{t}_{n}$. In the
continuum limit $\Delta\varphi$ becomes the angle of rotation of the normal $
\hat{n}$ and the binormal $\hat{b}$, with respect to the tangent $\hat{t}$
to the curve. In differential geometry this triad $\hat{t}$, $\hat{n}$ and $
\hat{b}$ is referred to as the Frenet frame and its rotation as one moves
along the curve is governed by the Frenet-Serret equations. As we will show
in the following, in the limit of a continuous curve the angle $\Delta\theta$
is proportional to the curvature (one can also show that the torsion is
proportional to the angle $\Delta\varphi$ \cite{shay}).

The calculation of statistical averages involves integration over all possible configurations of the polymer with respect to the measure $D\{\vec{r
}(s)\}$ or, alternatively, over $D\{\hat{t}(s)\}$. Discretization allows us
to replace this measure by the product of contributions of its segments, $D\{
\hat{t}(s)\}\rightarrow \Pi_n(d\hat{t}_n)$, where the measure of a single
segment is given by
\begin{equation}
d\hat{t}=\sin(\Delta\theta)d(\Delta\theta)d(\Delta\varphi)  \label{eq:Jaco_1}
\end{equation}
The local curvature of a discrete curve can be defined by replacing the
differentials by differences in Eq. \ref{eq:kappa}:
\begin{equation}
\Delta s\kappa_n=\sqrt{(\hat{t}_{n+1}-\hat{t}_n)^2}=\sqrt{
2(1-\cos(\Delta\theta))}  \label{eq:discreteKapp}
\end{equation}
where we used the equalities $(\hat{t}_{n+1})^2=(\hat{t}_n)^2=1$ and $\hat{t}
_{n+1}\cdot\hat{t}_n=cos(\Delta\theta)$. This definition of the curvature is
consistent with the one used in the Frenet-Serret theory only if $
\kappa\Delta s<<1$, in which case $\kappa\Delta s\approx|\Delta\theta|$.
Using Eqs.\ref{eq:Jaco_1} and \ref{eq:discreteKapp} it is straightforward
to show that:
\begin{equation}
d\hat{t}(s)=d(\Delta\varphi)(s)\Delta s^2\kappa(s) d\kappa(s)
\label{eq:Jaco_2}
\end{equation}
The above discussion applies to curves in 3 dimensions. In 2 dimensions the
measure is represented by a single angle $\theta$ so that in the discrete
representation of a curve, the angle between successive segments, $
\Delta\theta$, varies in the interval $[-\pi,\pi]$. Unlike the 3d case in
which the curvature is a positive definite quantity (see Eq. \ref{eq:kappa}
), the curvature of a planar curve can be either positive or negative (the
torsion is zero everywhere). The direction of the tangent vector can be
represented as a point on a unit circle and, therefore, the measure is given
by
\begin{equation}
d\hat{t}(s)=d(\Delta\theta)(s)\approx\Delta s d\kappa(s),\quad in \;2d
\label{eq:Jaco_2d}
\end{equation}
We conclude that in summing over the configurations of 3d curves, one can
replace the measure in the laboratory frame, $D\{\vec{r}(s)\}$, by that in
terms of the intrinsic coordinates of the deformed line, which is
proportional to $\kappa (s) D\{\kappa (s)\}$. This differs from the
corresponding measure for 2d curve which is proportional to $\propto
D\{\kappa (s)\}$.
\section{Curvature distribution of worm-like polymers}
In the worm-like chain model (WLC) a polymer is modeled as semi-flexible rod
whose elastic energy is governed only by the curvature \cite{Rubinstein}:
\begin{equation}
E_{WLC}=\frac{1}{2}b\int ds\kappa ^{2}(s)\rightarrow \frac{1}{2}
b\sum_{n}\Delta s\kappa _{n}^{2}  \label{eq:WLCenergy}
\end{equation}
where $b$ is the bending rigidity. This energy is invariant under polar
rotations and consequently the torsion angle $\Delta \varphi $ is uniformly
distributed in the range [$-\pi ,\pi $]. In 3 dimensions, the probability $
p(\kappa )d\kappa $ of observing a value of the curvature in the interval
between $\kappa (s)$ and $\kappa (s)+d\kappa $ at a point $s$ along the
contour of the curve, is proportional to the product of corresponding
Boltzmann factor by the measure $\kappa d\kappa $ (see Eq. \ref{eq:Jaco_2}),
\begin{equation}
p(\kappa )d\kappa =l_{p}\Delta se^{-\frac{1}{2}l_{p}\Delta s\kappa
^{2}}\kappa d\kappa   \label{eq:KappDistribution}
\end{equation}
where $l_{p}=b/k_{B}T$ is the persistence length (for $L>>l_{p}$ the polymer
length does not affect the statistics of the curvature). Inspection of Eq.\ref{eq:KappDistribution} shows that if one rescales the curvature as $\kappa \rightarrow \tilde{\kappa}=\kappa \sqrt{l_{p}\Delta s}$, the
probability distribution functions of polymers with different bending
rigidities can be represented by a single master curve $\tilde{\kappa}\exp
[-(1/2)\tilde{\kappa}^{2}]$. This scaling may lead to the conclusion that
changing the persistence length is equivalent to changing the
discretization. Indeed, calculation of properties such as the mean curvature
of a curve depends on the discretization length $\Delta s$ and diverges in
the continuum limit, $\left\langle \kappa \right\rangle =\lim_{\Delta
s\rightarrow 0}(l_{p}\Delta s)^{-1/2}=\infty $. This, however, is not true
in general and calculation of physical observables such as the
tangent-tangent correlation function, yields well-behaved results:
\begin{equation}
\left\langle \hat{t}(0)\cdot \hat{t}(s)\right\rangle =\left\langle \hat{t}_{0}\cdot \hat{t}_{n}\right\rangle =\left\langle \cos (\Delta \theta
)\right\rangle ^{n}  \label{eq:t0tn}
\end{equation}
From Eqs.\ref{eq:KappDistribution} and \ref{eq:discreteKapp} one can easily
conclude that: $\left\langle \cos (\Delta \theta )\right\rangle =1-\Delta
s/l_{p}\sim\exp [-\Delta s/l_{p}]$. Inserting the above expression
into Eq. \ref{eq:t0tn} one gets the WLC relation:
\begin{equation}
\left\langle \hat{t}(0)\cdot \hat{t}(s)\right\rangle =e^{-s/l_{p}}
\label{eq:t0ts}
\end{equation}
This correlation function and all properties that can be derived from it,
such as the mean square end to end distance, depend only on the persistence
length and are independent of the discretization.
\section{Simulations and analytical results}
\begin{figure}[p]
\centerline{\scalebox{0.6}{\includegraphics{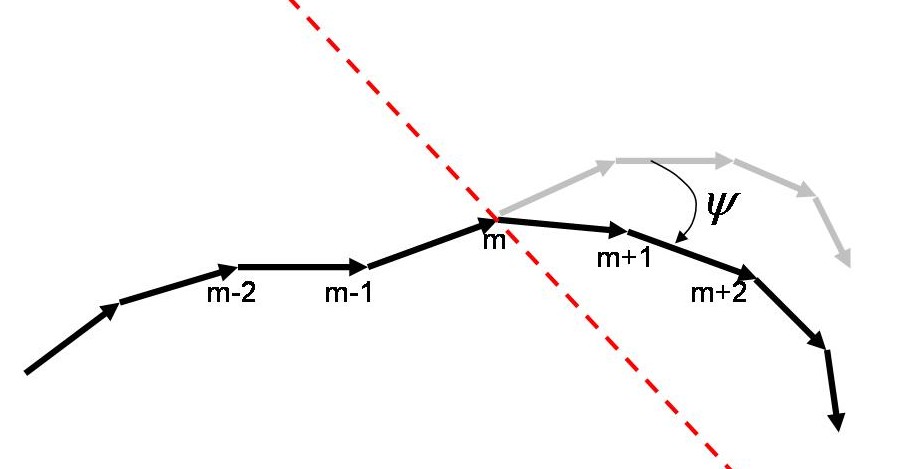}}}\caption{Caption of Fig. 2. }
\label{fig:model}
\end{figure}

\begin{figure}[ht]
\begin{center}
\includegraphics{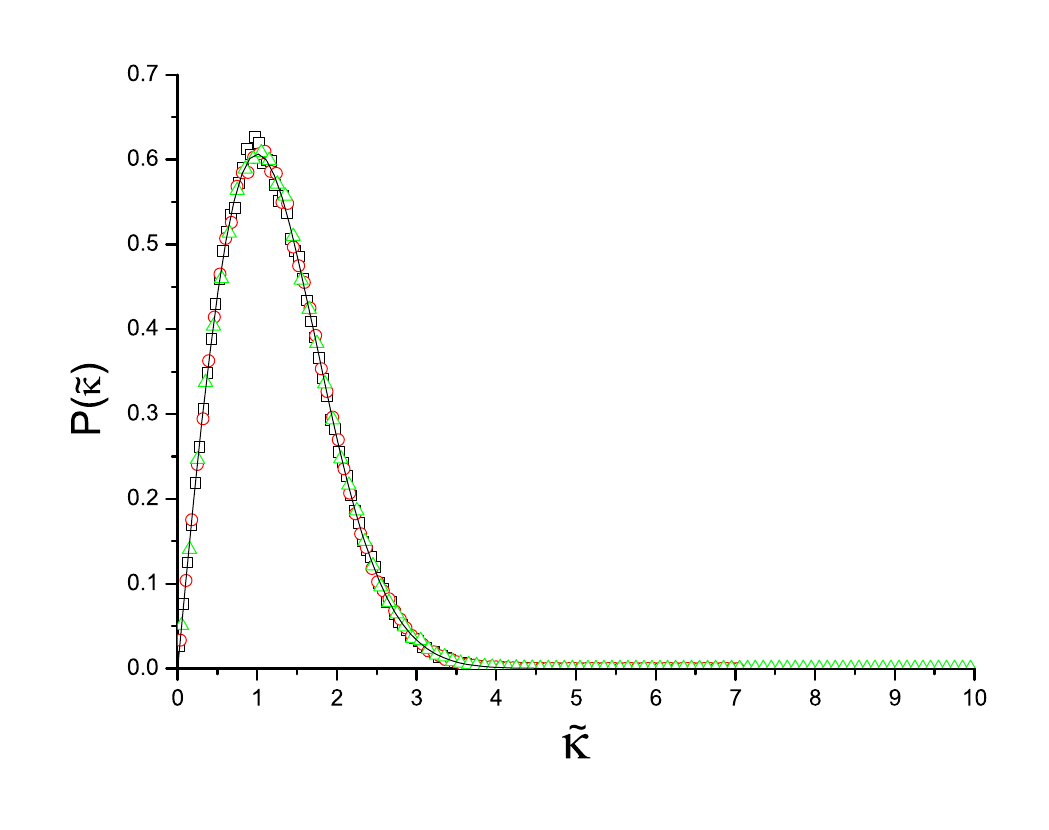}
\caption{Caption of Fig. 3.}
\label{fig:kappaDistribution}
\end{center}
\end{figure}

In order to study the curvature distribution function of a worm-like polymer
numerically, we performed Monte-Carlo simulations of a discrete polymer made
of N equal segments, in 3d. To generate the conformations of a polymer we
used the following moves (see ref. \cite{Jian}): a point between two
neighboring segments is picked at random and a randomly oriented axis that
passes through this point is defined. Then, the entire part of the polymer
that lies on one side of the chosen point, is rotated about the axis by an
angle $\psi $ which is uniformly distributed in the range $[-\pi ,\pi ]$
(see figure \ref{fig:model}). Since in the WLC model the total energy (in
dimensionless units, $k_{B}T=1$) of a discrete polymer is:
\begin{equation}
E_{WLC}=b\sum_{n}\left[ 1-\cos {\Delta \theta _{n}}\right] ,
\end{equation}
the change of energy in a move is given by
\begin{equation}
\Delta E_{WLC}=b\left[ \cos (\Delta \theta _{n})_{old}-\cos (\Delta \theta
_{n})_{new}\right]
\end{equation}
A move is accepted or rejected according to Metropolis rule and statistics
is collected following equilibration determined by convergence of the total
energy (after at least $10^{6}$ moves). In order to double check our method,
we calculated the tangent-tangent correlation function and the mean square
end to end distance as well and confirmed that they agree with well-known
analytical results for worm-like chains (not shown). Inspection of Fig. \ref
{fig:kappaDistribution} shows that there is perfect agreement between our
simulation results and the analytical expression, Eq. \ref
{eq:KappDistribution}. The two main features of the distribution are the
existence of a peak at a finite value of the curvature and the fact that the
probability to observe zero curvature vanishes identically. At first sight,
the prediction that any local measurement on a worm-like chain will yield a
non-vanishing curvature, contradicts our intuition about stiff polymers.
Note, however, that experiments (e.g., by AFM) that monitor the
configurations of semi-flexible polymers such as dsDNA, do not directly
measure the curvature but rather the local bending angle. Even though $
\left\langle \kappa \right\rangle =1/\sqrt{l_{p}\Delta s}$ does not vanish,
the average local angle (measured on length scales smaller than the
persistence length) $\left\langle \delta \theta \right\rangle =\sqrt{\Delta
s/l_{p}}$ is always small.

\begin{figure}[ht]
\centerline{(a)\includegraphics[width=0.5\linewidth]{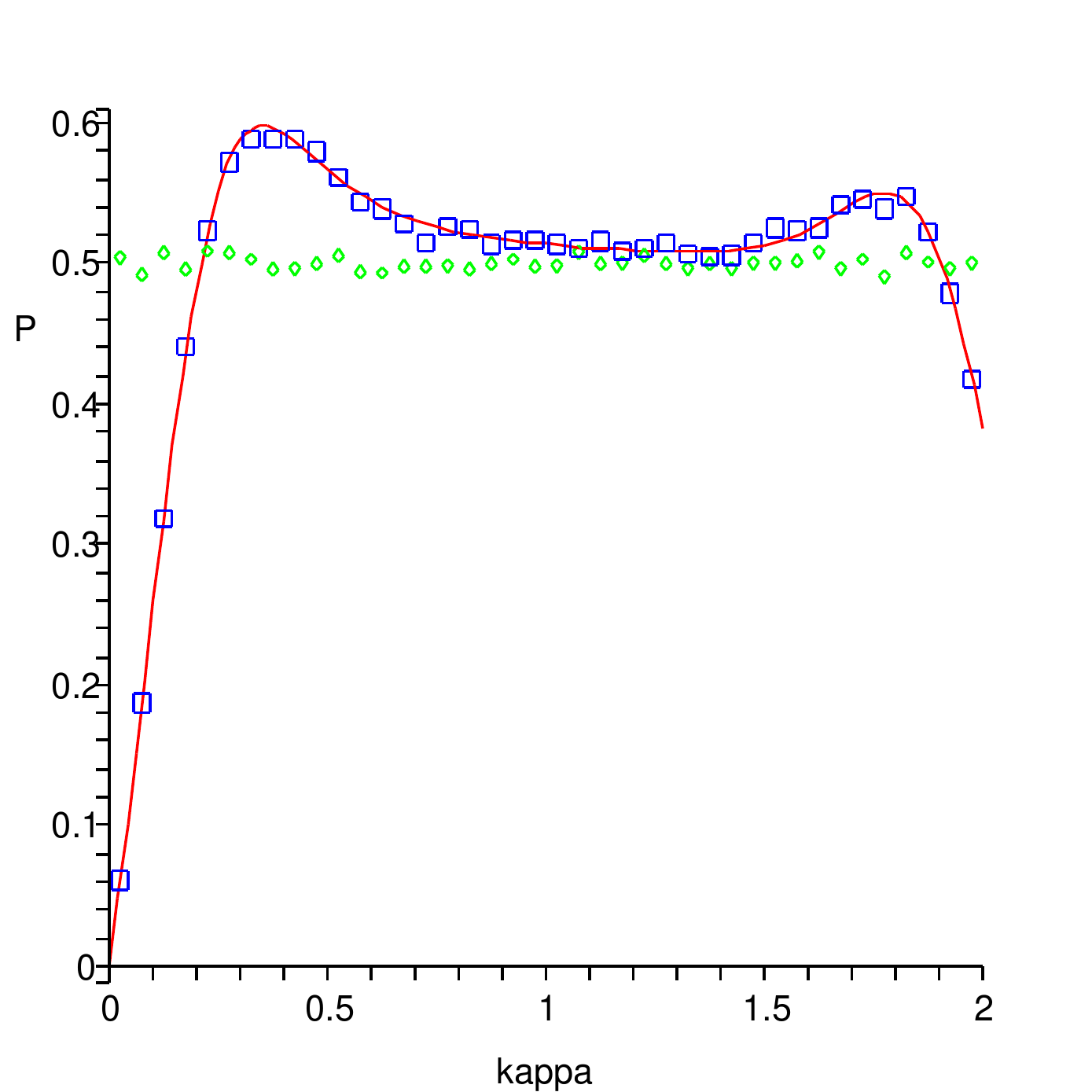}
(b)\includegraphics[width=0.5\linewidth]{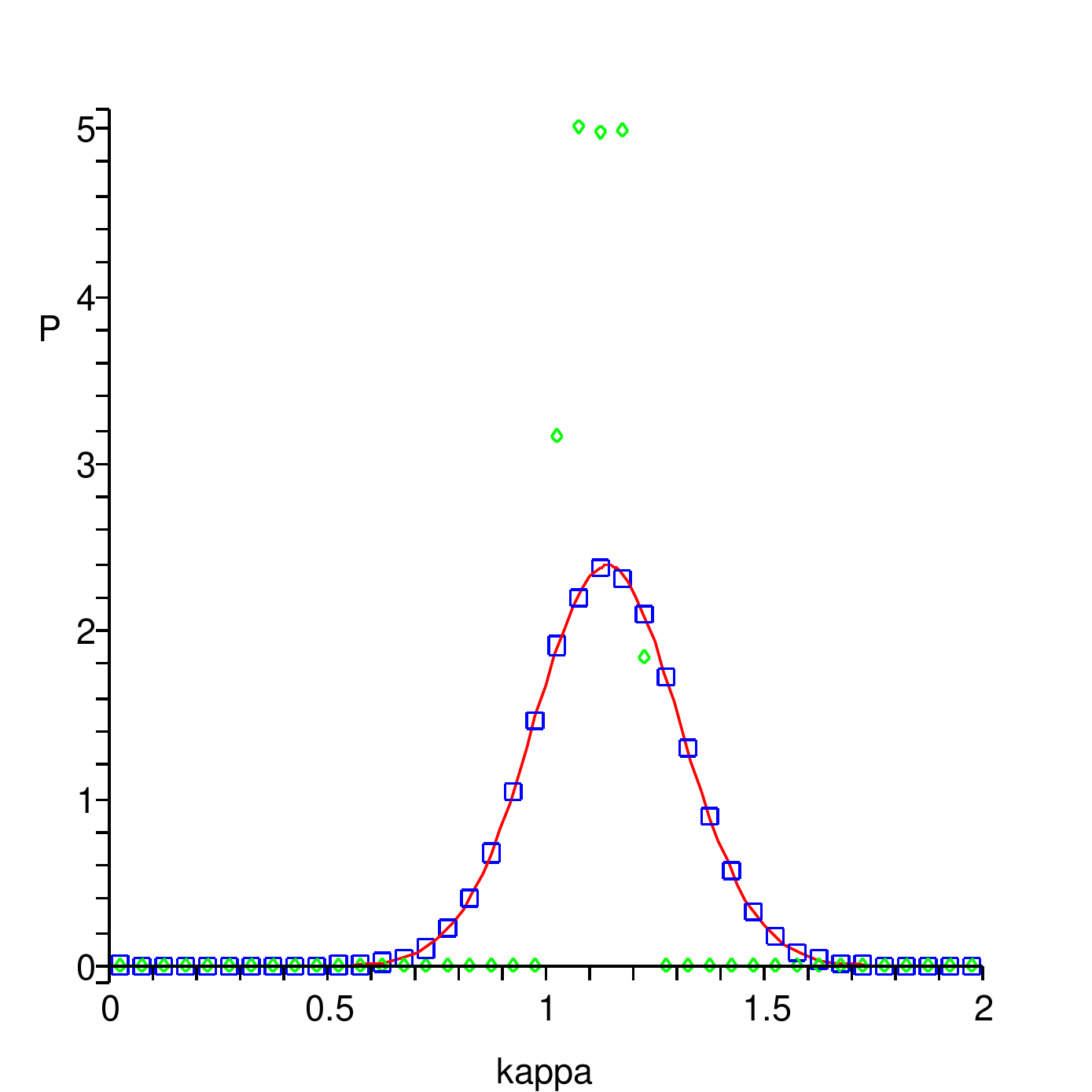}}
\centerline{(c)\includegraphics[width=0.5\linewidth]{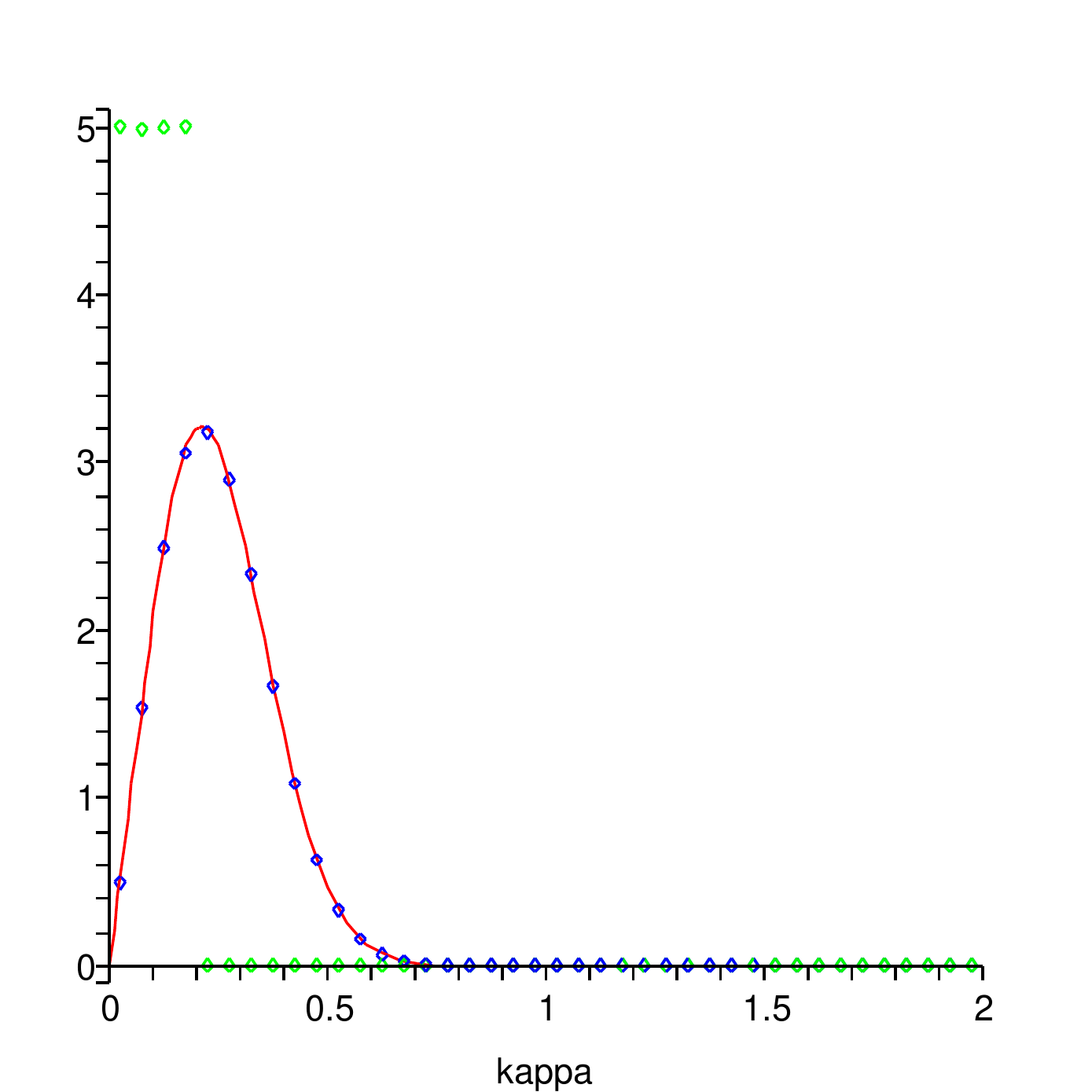}
(d)\includegraphics[width=0.5\linewidth]{Pk_40_2.pdf}}
\caption{Caption of Fig. 4.}
\label{fig:intrinsicCurvature}
\end{figure}

The above discussion applies to worm-like polymers that do not have
intrinsic curvature (their curvature is generated by thermal fluctuations
only). Although the WLC model has been applied \cite{Marko-Siggia} to
interpret mechanical experiments on individual double stranded DNA (dsDNA)
molecules \cite{Bustemante-Bensimon}, it has been suggested\cite{trifonov}
that dsDNA has sequence-dependent intrinsic curvature. In this case, the
experimentally observed local curvature of dsDNA contains both the
contribution of intrinsic curvature and that of thermal fluctuations \cite
{Scipioni}. The generalization to the case of spontaneous curvature is
straightforward: even though the expression for the bending energy Eq. \ref
{eq:WLCenergy}, is replaced by
\begin{equation}
E_{WLC}=\frac{1}{2}b\int ds\delta \kappa ^{2}(s)\rightarrow \frac{1}{2}
b\sum_{n}\Delta s(\kappa _{n}-\kappa _{0,n})^{2}  \label{eq:DNAWLCenergy}
\end{equation}
the measure is not affected and depends only on the dimensionality of the
problem. Using the discrete form of curvature Eq. \ref{eq:discreteKapp} the
distribution function of curvature for a given intrinsic curvature $\tilde{
\kappa}_{0}$ is:
\begin{equation}
p(\tilde{\kappa}|\tilde{\kappa}_{0})d\tilde{\kappa}=A(\tilde{\kappa}_{0})
\tilde{\kappa}e^{-\frac{1}{2}l(\tilde{\kappa}-\tilde{\kappa}_{0})^{2}}d
\tilde{\kappa}  \label{eq:Pkk0}
\end{equation}
where
\begin{equation}
A(\tilde{\kappa}_{0})=\left[ \int d\tilde{\kappa}\tilde{\kappa}\exp (-\frac{1
}{2}l(\tilde{\kappa}-\tilde{\kappa}_{0})^{2})\right] ^{-1/2}
\end{equation}
is a normalization factor and $l$ is the persistence length. An experiment
measuring the curvature distribution of an ensemble of conformations of a
polymer with a specific primary sequence $\left\{ \kappa _{0}(s)\right\} $,
would measure the distribution Eq. \ref{eq:Pkk0}, from which the intrinsic
curvature can be calculated as a fit parameter. If one monitors the
curvature distribution of an ensemble of random copolymers whose sequences
are characterized by a distribution of intrinsic curvature $p(\tilde{\kappa}
_{0}),$ the corresponding curvature distribution function will be:
\begin{equation}
p(\tilde{\kappa})=\int d\tilde{\kappa}_{0}p(\tilde{\kappa}|\tilde{\kappa}
_{0})p(\tilde{\kappa}_{0})
\label{eq:kapDistribution}
\end{equation}
In order to check this result we performed Monte Carlo simulations with different distributions
of $\tilde{\kappa}_{0}$ . As can be observed in Fig:\ref{fig:intrinsicCurvature}, the simulations are
in perfect agreement with the analytical results for the curvature distribution calculated
using Eq. \ref{eq:kapDistribution}
\section{Discussion}
Finally, we would like to comment on the applicability of our results to present
and future experiments. To the best of our knowledge,
all experiments that probed the local curvature of DNA (by electron
microscopy \cite{Muzard-Bednar} and by SFM \cite{Rivetti,Scipioni}) were
done on dsDNA molecules deposited on a surface. It has been argued that in
this case one has to distinguish between strong adsorption in which case
dsDNA attains a purely 2d conformation, and weak adsorption in which the 3d
character of the adsorbed macromolecule is maintained, at least in part (see
ref. \cite{Joanicot}). While in the former case the measure of Eq. \ref
{eq:Jaco_2d} applies and the most probable curvature vanishes \cite{Cognet},
in the latter case one should use the 3d measure of Eq. \ref{eq:Jaco_2} for
which the most probable curvature is finite and the probability to observe
zero curvature vanishes. New experiments that could probe the 3d
conformations of DNA and other worm-like polymers are clearly necessary to
test our predictions. 
\\

{\bf Acknowledgments} 
We would like to acknowledge correspondence with A.Yu. Grosberg
Y. Rabin would like to acknowledge
support by a grant from the US-Israel Binational Science Foundation.

\clearpage
\begin{list}{}{\leftmargin 2cm \labelwidth 1.5cm \labelsep 0.5cm}

\item[\bf Fig. 1] 3 successive segments are shown by black arrows. $\Delta\protect
\theta$ is the angle between the $(n-1)$'th segment (the radius of the unit
sphere) and the $n$'th segment. $\Delta\protect\varphi$ is the polar angle
defined by the projection of $\hat{t}_n$ (dashed arrow) on the plane defined
by the normal $\hat{n}_{n-1}$ and the binormal $\hat{b}_{n-1}$ (red arrows)
\item[\bf Fig. 2] Sketch of a Monte Carlo move. The right hand part of the curve rotates with random angle $\psi$ with respect to a random axis (dashed red line) passing through a random point $m$
\item[\bf Fig. 3] Distribution function of scaled curvature $\tilde{\kappa}$ for a polymer of $N=10^4$ segments of length $\Delta s=1$ each. The empty symbols are the simulation results for $l_p=25$ (black squares),  $l_p=50$ (red circles) and $l_p=100$ (green triangles). The rigid line is the analytical scaled distribution function $\tilde{\kappa}\exp[-(1/2)\tilde{\kappa}^2]$ where $\tilde{\kappa}=\kappa\sqrt{l_p\Delta s}$. There are no fit parameters.
\item[\bf Fig. 4] In all figures the blue squares and green diamonds are the histograms of curvature and of intrinsic curvature, respectively. The red line is the analytical distribution function computed using Eq. \ref{eq:kapDistribution}.

\end{list}

\clearpage

\end{document}